\RequirePackage{fix-cm}
\documentclass[twocolumn,epjc3]{svjour3}  
\smartqed  
\RequirePackage{graphicx}
\journalname{Eur. Phys. J. C}
\usepackage{amsmath}
\usepackage{cite}
\usepackage[colorinlistoftodos]{todonotes}
\usepackage[colorlinks=true, allcolors=blue]{hyperref}

\begin{document}

\title{Stability of cosmic structures in scalar-tensor theories of gravity
}


\author{Grigoris Panotopoulos \thanksref{e1,addr1}
        \and
        {\'A}ngel Rinc{\'o}n \thanksref{e2,addr2} 
}

\thankstext{e1}{e-mail: \href{mailto:grigorios.panotopoulos@tecnico.ulisboa.pt}{\nolinkurl{grigorios.panotopoulos@tecnico.ulisboa.pt}}}
\thankstext{e2}{e-mail: \href{mailto:arrincon@uc.cl}{\nolinkurl{arrincon@uc.cl}}}


\institute{Centro Multidisciplinar de Astrof{\'i}sica, Instituto Superior T{\'e}cnico, 
\\
Universidade de Lisboa, Av. Rovisco Pais, 1049-001 Lisboa, Portugal \label{addr1}
           \and
           Instituto de F{\'i}sica, Pontificia Universidad Cat{\'o}lica de Chile,\\ Av. Vicu{\~n}a Mackenna 4860, Santiago, Chile \label{addr2}
}

\date{Received: date / Accepted: date}

\maketitle

\begin{abstract}
In the present work we study a concrete model of Scalar--Tensor theory of gravity characterized by two free parameters, and compare its predictions against observational data and constraints coming from supernovae, solar system tests as well as the stability of cosmic structures. First an exact analytical solution at the background level is obtained. Then using that solution the expression for the turnaround radius is computed. Finally we show graphically how current data and limits put bounds on the parameters of the model at hand.

\end{abstract}

\section{Introduction}

In the end of the 90's the most dramatic discoveries in Particle Physics and Cosmology were from the one hand the neutrino oscillations and on the other hand the current acceleration of the Universe \cite{SN1,SN2}. Nowadays many well established observational data from Astrophysics and Cosmology show that we live in a spatially flat Universe that expands in an accelerating rate \cite{turner}. Dark energy, the fluid component that dominates the evolution of the Universe and drives the current cosmic acceleration, is one of the biggest challenges of modern cosmology, as its nature and origin still remains a mystery. The $\Lambda$CDM model with a constant equation of state $w=-1$ is the most economical one in excellent agreement with current data. However, given the cosmological constant problems other alternatives with an evolving equation of state have been studied in the literature over the years. In general all
dark energy models fall into two broad classes, namely from the one hand dynamical dark energy models in which one has to introduce a new dynamical field assuming Einstein's General Relativity (GR) \cite{quint,mukhanov,tachyon,chapl}, and from the other hand geometrical dark energy models in which one assumes an alternative theory of gravity that modifies GR at cosmological scales. In the latter category we find the well-known examples of $f(R)$
theories of gravity \cite{Sotiriou:2008rp,DeFelice:2010aj,Nojiri:2017ncd,Nojiri:2010wj}, the Dvali-Gabadadze-Porrati brane model \cite{dgp} or Scalar-Tensor theories of gravity (ST), with the Brans-Dicke \cite{BD} model being the archetypical one and recently the scale dependent approach previously applied to certain black holes problems \cite{Koch:2016uso,Rincon:2017ypd,Rincon:2017goj,Rincon:2017ayr,Contreras:2017eza}.

Until a few years ago the observational data used to constrain dark energy models were mainly the temperature anisotropies of the cosmic microwave background, galaxy surveys and supernovae data. However, recently a new potentially local check was proposed in \cite{tomaras} based on two facts, namely a) the motion of a test particle depends on the interplay between the initial momentum of the Big-Bang, the attractive nature of gravity and the repulsive nature of dark energy, and b) for a given mass of a spherical structure there is a maximum radius, called the turnaround point $R_T$, beyond which a test particle cannot stay bound due to the antigravity effect of dark energy. This is very similar to what happens in neutron stars where the mass-to-radius relation depends crucially on the poorly known equation-of-state \cite{eos}, and observed pulsars with a mass at two solar masses rule out equations-of-state that predict a lower higher value for the star mass \cite{shapiro}. Then in \cite{tomaras1} the authors considered dark energy models with a constant equation-of-state $w$, and soon after that the idea was further pursued in subsequent works applied to generic dark anergy models \cite{tomaras2}, Brans-Dicke theory \cite{tomaras3} and DGP brane model \cite{kousvos}. Unfortunately, in novel cosmologies characterized by non-standard Friedmann-like equations it is highly non-trivial to see the implications of the prediction of the models regarding the maximum turnaround point, although a formula for $R_T$ may exist.

ST theories of gravity are straight-forward generalizations and in fact the simplest extension of GR. Given that the stability of cosmic structures based on the maximum turnaround point has not adequately analyzed yet, in the present work we ask ourselves the question what the stability of cosmic structures together with other observational data and limits can tell us about cosmologies based on ST theories of gravity. The goal of this work is two-fold. First we present an exact analytical cosmological solution of a ST theory of gravity, which is always desirable,
and then we compute the turnaround radius of the model. Our work is organized as follows: After this introduction we present the model, the cosmological equations at the background level as well as the exact analytical solution in section two. In the third section we compute the expression for the turnaround radius, and in section \ref{data_comparison} we use current data to constrain the parameters of the model. Finally we conclude our work in the fifth section. We use natural units such that $c = 8 \pi G = \hbar = 1$ and metric signature $(-, +, +,+)$. 

\section{The model, the cosmological equations and the exact solution}
We start by defining the model
\begin{equation}
\begin{split}
S[g_{\mu \nu},\phi] = \frac{1}{2} \int \mathrm{d}^4x \sqrt{-g} \Bigl[& F(\phi) R -g^{\mu \nu} \partial_\mu \phi \partial_\nu \phi 
\\
& - 2 V(\phi)\Bigl] \ + \ S_m
\end{split}
\end{equation}
where $S_m$ is the action of matter fields, $g$ is the determinant of the metric $g_{\mu \nu}$, $R$ is the Ricci scalar, $\phi$ is the scalar field and $V(\phi)$ is the scalar self-interaction potential. The dimensionless function $F(\phi)$ describes the variation of the effective gravitational constant. This is a generalization of quintessence models and it is characterized not only by the scalar potential but also by $F(\phi)$. Considering a flat FRW ansatz for the metric
\begin{equation}
ds^2 = -dt^2 + a(t)^2 [d r^2 + r^2 (d \theta^2 + \sin^2 \theta d \varphi^2)]
\end{equation}
with $t$ being the cosmic time and $a(t)$ being the scale factor, and assuming no interaction between the scalar field and the non-relativistic matter
with pressure $p=0$ and energy density $\rho$, one obtains the following cosmological equation for the background \cite{leandros1,leandros2}
\begin{eqnarray}
3 F H^2 & = & \rho + \frac{1}{2} \dot{\phi}^2 + V - 3 H \dot{F} \\
-2 F \dot{H} & = & \rho + \dot{\phi}^2 + \ddot{F} - H \dot{F} \\
\dot{\rho} + 3 H \rho & = & 0 \\
\ddot{\phi} + 3 H \dot{\phi} & = & 3 F_{,\phi} (\dot{H}+H^2) - V_{,\phi}
\end{eqnarray}
where $H=\dot{a}/a$ is the Hubble parameter, the dot denotes differentiation with respect to cosmic time, while the $,\phi$ denotes differentiation with respect to the scalar field. 
Clearly, when $F(\phi)=1$ we recover the standard equations valid in GR.
Note that there are four equations in total, but only three of them are independent. 
In addition, power-law solutions are very common in cosmological models based on GR in various contexts.
Besides the trivial examples of the radiation and the matter dominated era, one can mention the
well-known cases of the power-law inflation \cite{Lucchin:1984yf} as well as the study of
cosmological scaling solutions \cite{Liddle:1998xm}.
Thus,
we seek power-law solution of the form
\begin{eqnarray}
F(\phi) & = & \lambda \left(\frac{\phi}{\phi_0}\right)^2 \\
V(\phi) & = & V_0 \left(\frac{\phi}{\phi_0}\right)^\alpha \\
a(t) & = & a_0\left(\frac{t}{t_0}\right)^p \\
\phi(t) & = & \phi_0 \left(\frac{t_0}{t}\right)^m
\end{eqnarray}
where the subindex 0 denotes present values and $a_0$ is defined at the present time as unity, and $p > 1$ corresponds to accelerating solutions. Plugging everything into the equations
one can check that all of them are satisfied provided that
\begin{eqnarray}
\alpha & = & \frac{6p}{3p-2} \\
m & = & \frac{3}{2}p - 1 \\
\phi_0^2 & = & \frac{8}{(3p-2)^2} \bigg[V_0^2 t_0^2-\lambda (6p^2-7p+2) \bigg] \\
3 \lambda p^2 & = & (V_0+\rho_0) t_0^2+3 p \lambda (3p-2) + \frac{1}{8}\phi_0^2 (3p-2)^2 \\
8 p \lambda & = & 4 \rho_0 t_0^2 + \phi_0^2 (3p-2)^2 + 4 \lambda (12p^2-11p+2)
\end{eqnarray}
and it is easy to verify that combining any two of the last three equation we obtain the third. Therefore, we can choose $p,\lambda$ to be the free parameters of the model, while $\phi_0,V_0$ are determined by the previous expressions. This is our first main result in this work. We remark in passing that exact analytical solutions have been obtained in \cite{exact}, but without matter. 

\begin{figure}[ht!]
\centering
\includegraphics[width=\linewidth]{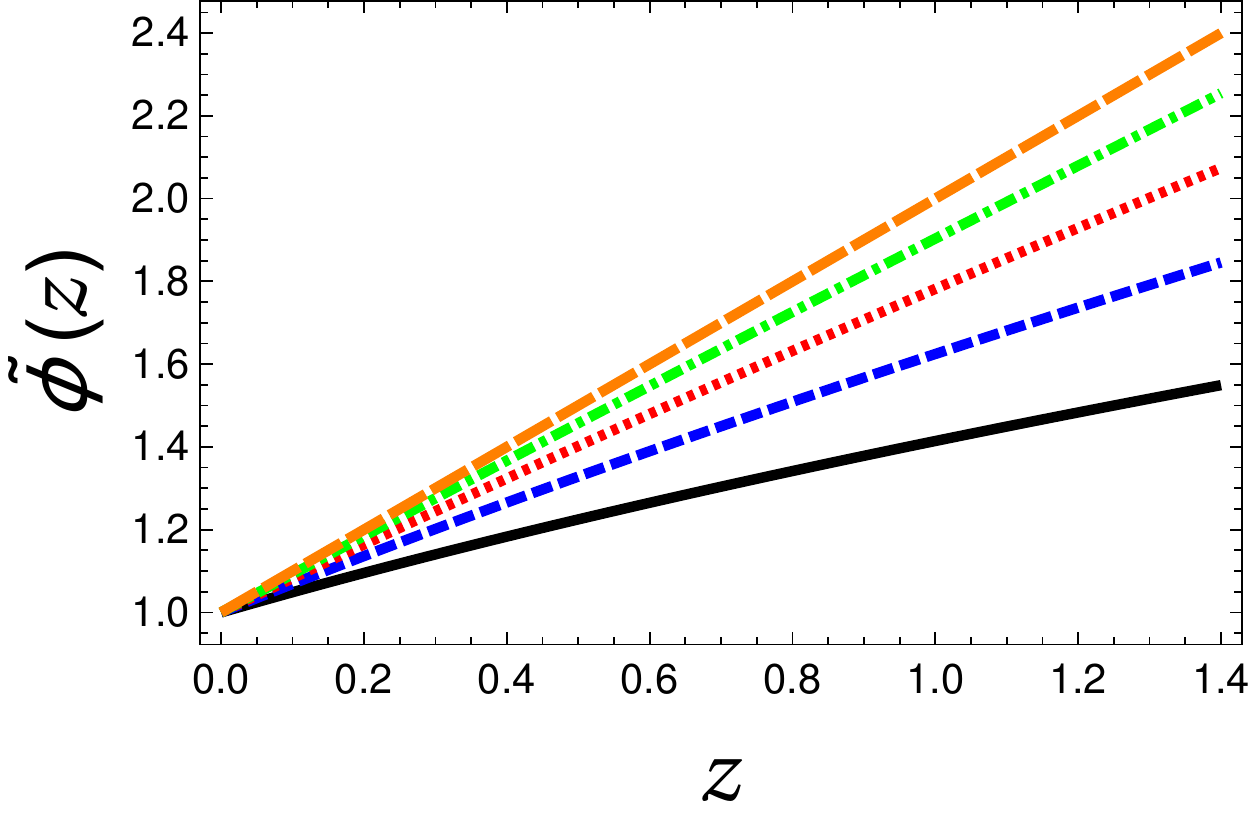}
\caption{\label{fig:1} 
Dimensionless function $\tilde{\phi}(z)$ versus red--shift $z$ for different values of the parameter $p$. The curves correspond to: $p=1$ (solid black line), $p=1.25$ (short dashed blue line), $p=1.5$ (dotted red line), $p=1.75$ (dotted dashed green line) and $p=2$ (long dashed orange line).
}
\end{figure}

\begin{figure}[ht!]
\centering
\includegraphics[width=\linewidth]{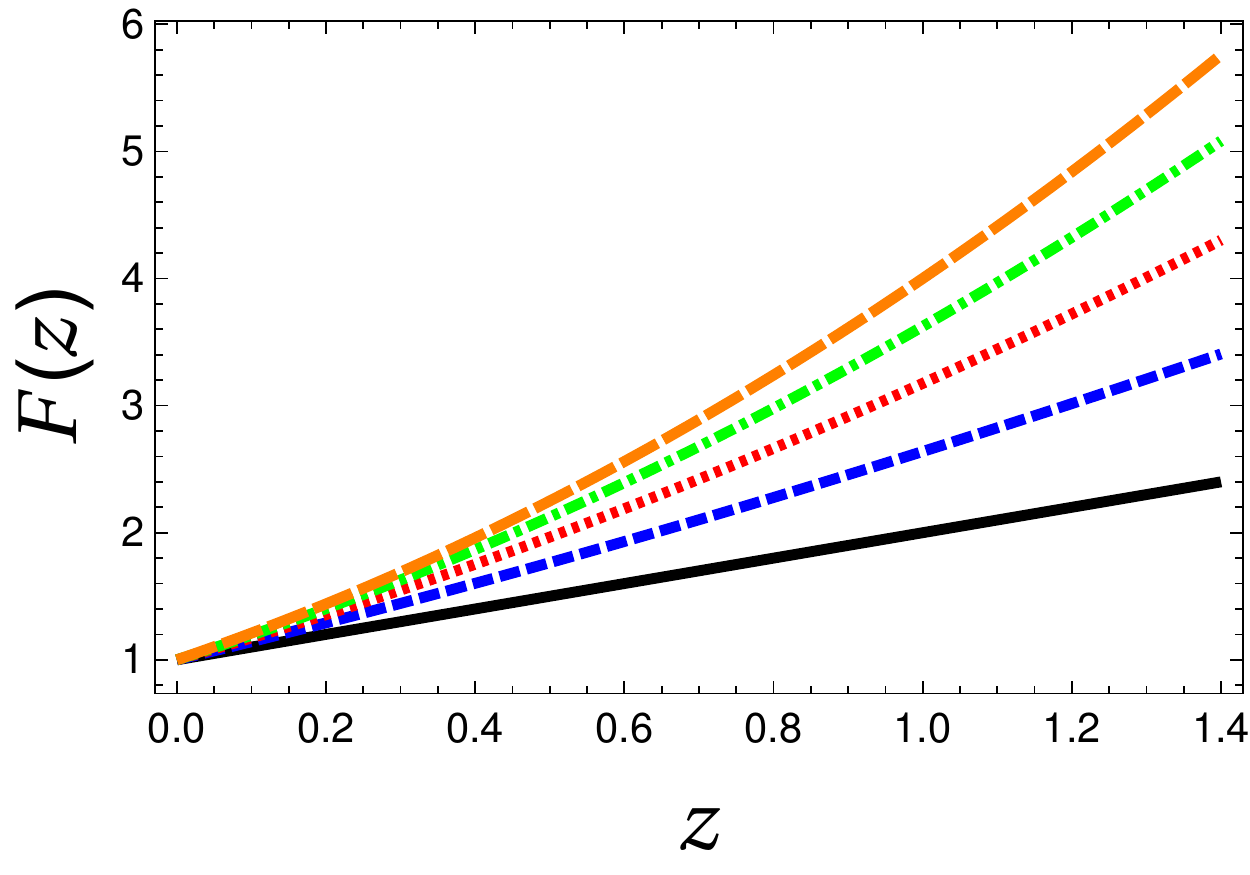}
\caption{\label{fig:2} 
Dimensionless function $F(z)$ versus red--shift $z$ for different values of the parameter $p$. The curves correspond to: $p=1$ (solid black line), $p=1.25$ (short dashed blue line), $p=1.5$ (dotted red line), $p=1.75$ (dotted dashed green line) and $p=2$ (long dashed orange line). Note that the vertical axis is scaled to $1 : 10^{-2}$.
}
\end{figure}

\begin{figure}[ht!]
\centering
\includegraphics[width=\linewidth]{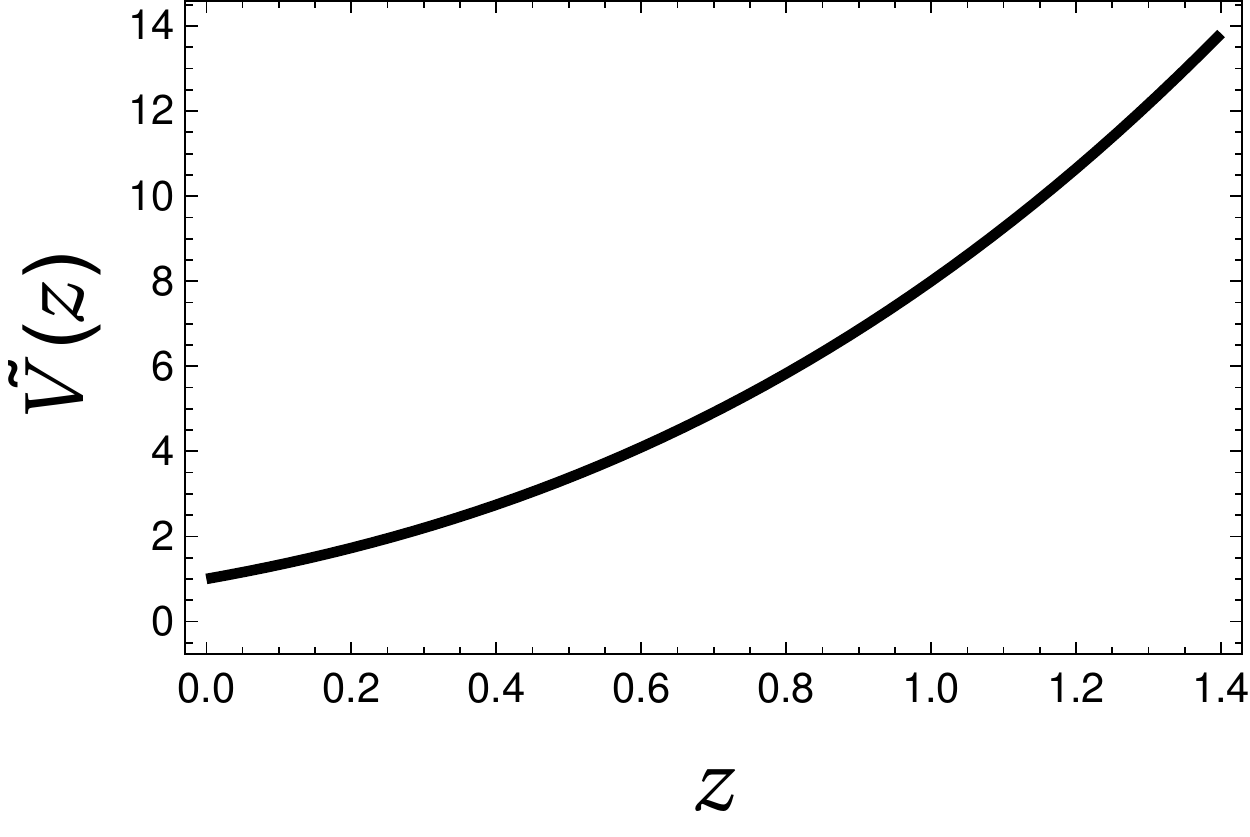}
\caption{\label{fig:3} 
Dimensionless function $\tilde{V}(z)$ versus red--shift $z$.
}
\end{figure}

Next the behaviour of the set of fields $\{\phi, F, V\}$ as functions of red--shift $z=-1+1/a$ is investigated. We combine the aforementioned fields to obtain the dimensionless functions involved, namely
\begin{align}
\tilde{\phi}(z)  & = \hspace{0.25cm} (1+z)^{\frac{3}{2}-\frac{1}{p}}
\\
F(z) &= \lambda (1+z)^{3-\frac{2}{p}}
\\
\tilde{V}(z) & =  \hspace{0.25cm} (1+z)^3
\end{align}
where $\tilde{\phi}(z) \equiv \phi(z)/\phi_0 $, $\tilde{V}(z) \equiv V(z)/V_0$, whereas $F(z)$ is dimensionless by definition.  Note that in our scalar-tensor model, the potential $V(z)$ does not depend on the free parameter $p$. We plot $\tilde{\phi}(z)$, $F(z)$ and $\tilde{V}(z)$ as functions of red--shift for different values of $p$. The three quantities are shown in Figures \ref{fig:1},  \ref{fig:2} and  \ref{fig:3} respectively.

\section{The maximum turnaround radius in the ST model}

To study cosmic structures we need to study the evolution of the metric scalar perturbations $\Psi(\eta,x^i), \Phi(\eta,x^i)$ defined by \cite{dePutter:2010vy,Albarran:2016mdu,Panotopoulos}
\begin{equation}
ds^2 = a(\eta)^2 [-(1+2 \Phi) d\eta^2 + (1-2 \Psi) \delta_{i j} dx^i dx^j]
\end{equation}
with $d\eta=dt/a$ being the conformal time. We also need the conservation equation for the peculiar velocity $\delta u^i$ \cite{dePutter:2010vy,Albarran:2016mdu,Panotopoulos}
\begin{equation}
u'^i + \mathcal{H} \delta u^i = - \partial_i \Phi
\end{equation}
where $\mathcal{H}$ is the conformal Hubble parameter.
Following \cite{kousvos} we consider a shell of backreactionless 
cold dark matter fluid moving just outside the structure. The physical spatial coordinate corresponding to the cold dark matter perturbation is 
\begin{equation}
r^i = a(\eta) x^i
\end{equation}
We can now obtain the velocity as well as the acceleration of this element as follows:
First the velocity is computed to be 
\begin{equation}
\frac{d r^i}{d t} = \frac{1}{a(\eta)} \frac{d r^i}{d \eta} = \delta u^i + \mathcal{H} x^i
\end{equation}
while taking the derivative once more we obtain the acceleration 
\begin{equation}
\frac{d^2 r^i}{d t^2} = \frac{\mathcal{H}'}{a^2} r^i \partial_i \Phi
\end{equation}
In a non-standard cosmology the Poisson equation for sub-horizon scales becomes \cite{kousvos}
\begin{equation}
\nabla^2{\Phi} = 4 \pi G_{\text{eff}} \delta \rho
\end{equation}
where $\delta \rho$ is the perturbation of the matter energy density, and $G_{\text{eff}}$ is the effective Newton's constant which is different than $G_N$. In the last step we approximate the whole structure as a point mass located at $\vec{r}=\vec{0}$, and the source in the Poisson equation reads
\begin{equation}
\delta \rho = \mathcal{M} \delta^{3} (\vec{r})
\end{equation}
with $\mathcal{M}$ being the mass of the structure. Then the Poisson equation becomes
\begin{equation}
\nabla^2{\Phi} = 4 \pi G_{\text{eff}} \mathcal{M} \delta^{3} (a(\eta) \vec{R})
\end{equation}
and therefore the solution reads
\begin{equation}
\Phi = - \frac{G_{\text{eff}} \mathcal{M}}{R}
\end{equation}
The maximum turnaround radius by definition is computed by requiring that the acceleration vanishes at that point. Thus we obtain \cite{kousvos}
\begin{equation}
\dot{H} R_T + \frac{G_{\text{eff}} \mathcal{M}}{R_T^2} = 0
\end{equation}
which implies
\begin{align}
R_T = \left[\frac{G_{\text{eff}}}{|\dot{H}|}\mathcal{M}\right]^{1/3}
\end{align}
We see that the maximum turnaround point depends on the interplay between the background evolution $\dot{H}$ and the effective Newton's constant $G_{\text{eff}}$. 
In Scalar-Tensor theories of gravity the effective Newton's constant is given by \cite{leandros1,leandros2}
\begin{equation}
\frac{G_{\text{eff}}}{G_N} = \frac{1}{F} \left[ \frac{2F + 4 F_{,\phi}^2}{2F + 3 F_{,\phi}^2} \right] \simeq \frac{1}{F}
\end{equation}
assuming that $F \gg F_{,\phi}^2$ (which in the end we check that it indeed holds), while constraints from solar system tests require that \cite{leandros1}
\begin{equation}
\omega_0^{-1} = F_{,\phi}|_0^2 < 4 \times 10^{-4}
\end{equation}
Given the exact solution we obtained in the previous section we can now compute both $\dot{H}$ and $G_{\text{eff}}$, and we finally obtain for $R_T$ the expression
\begin{equation}
R_T = \left ( \frac{p}{\lambda} \frac{G_N \mathcal{M}}{H_0^2}  \right)^{1/3}
\end{equation}
where we have evaluated all the time-dependent quantities at today since the cosmic structures we consider here are nearby structures and thus they correspond to $z \simeq 0$. This is our second main result in the present article.

We recall at this point that in \cite{tomaras1,tomaras2} it was found that for the $\Lambda$CDM model, where $w=-1$, the turnaround point is given by
\begin{equation}
R_{T,st} = \left ( \frac{G_N \mathcal{M}}{\Omega_{\Lambda,0} H_0^2}  \right)^{1/3}
\end{equation}
Therefore to compare with the $\Lambda$CDM model we write the previous formula equivalently as follows
\begin{equation}
R_{T} = R_{T,st} \left ( \frac{p \Omega_{\Lambda,0}}{\lambda} \right)^{1/3}
\end{equation}
Therefore we see that the ST cosmological model studied here agrees with the $\Lambda$CDM model when the ratio $x = p/\lambda \sim 1$. 

\section{Comparison of the model with data}\label{data_comparison}

Finally, in this section we briefly compare the ST model considered here against observational data from a) supernovae data, b) solar system tests, and c) stability of cosmic structures based on the maximum turnaround radius obtained in the previous section.


\subsection{Supernovae data}

\begin{figure}[ht!]
\centering
\includegraphics[width=\linewidth]{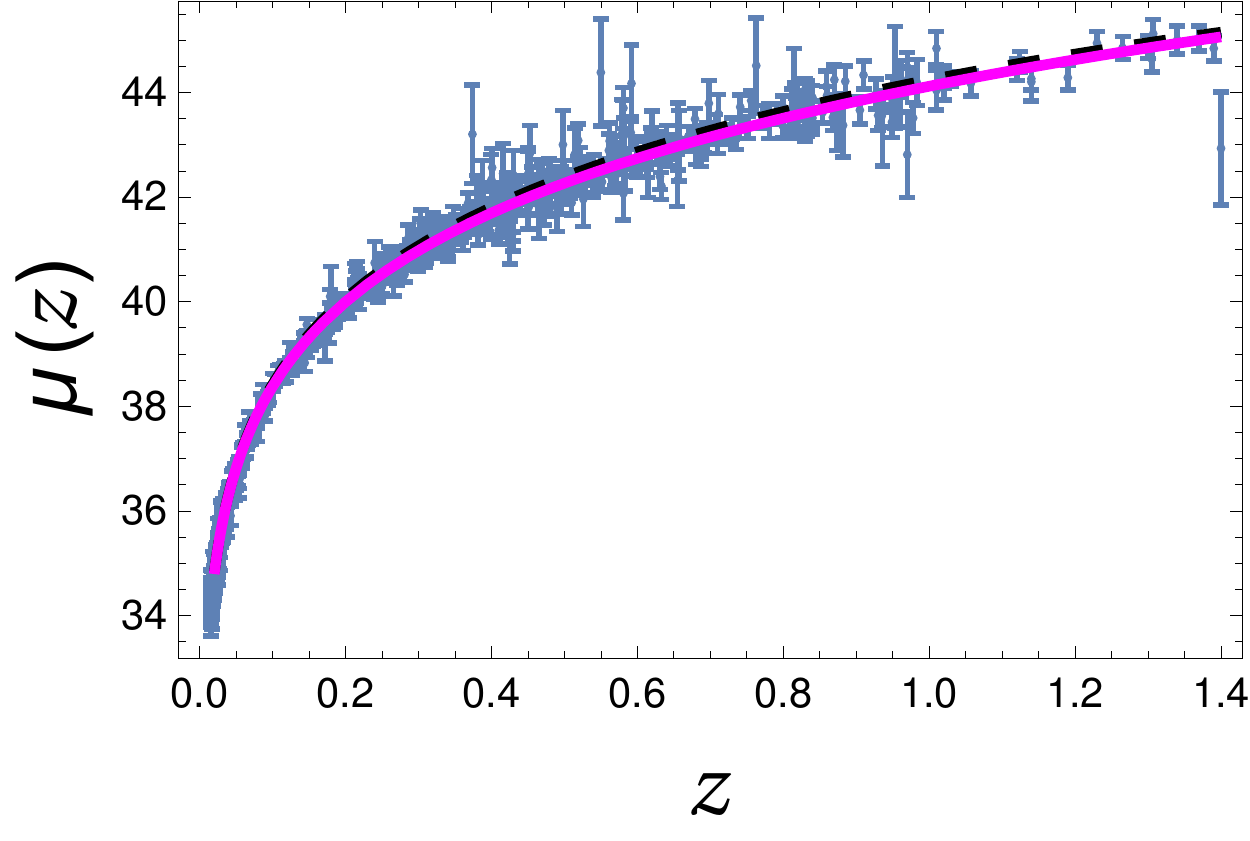}
\caption{\label{fig:4} 
Distance modulus versus red-shift for $\Lambda$CDM (dashed curve in black) and for the ST model for $p=1.25$ (solid curve in magenta). The supernovae data are from the Union 2 compilation.
}
\end{figure}

The Hubble parameter as a function of the red-shift $z=-1+1/a$ is computed to be
\begin{equation}
H(z) = H_0 (1+z)^{\frac{1}{p}}
\end{equation}
while the luminosity distance is given by \cite{hogg}
\begin{equation}
d_L(z) = (1+z) \int_0^z dx \frac{1}{H(x)} 
\end{equation}
Finally the supernovae distance modulus $\mu = m-M$, where $M$ is the absolute and $m$ the apparent magnitude, is given by \cite{copeland,nesseris}
\begin{equation}
\mu(z) = 25 + 5 \log_{10} \left[ \frac{d_L(z)}{\text{Mpc}} \right]
\end{equation}
In Fig. \ref{fig:4} we show the distance modulus as a function of the red-shift both
for $\Lambda$CDM and for the ST model studied here for $p=1.25$. Observational data from the Union 2 compilation \cite{union2} are shown too.


\subsection{Solar system tests}
\begin{figure}[ht!]
\centering
\includegraphics[width=\linewidth]{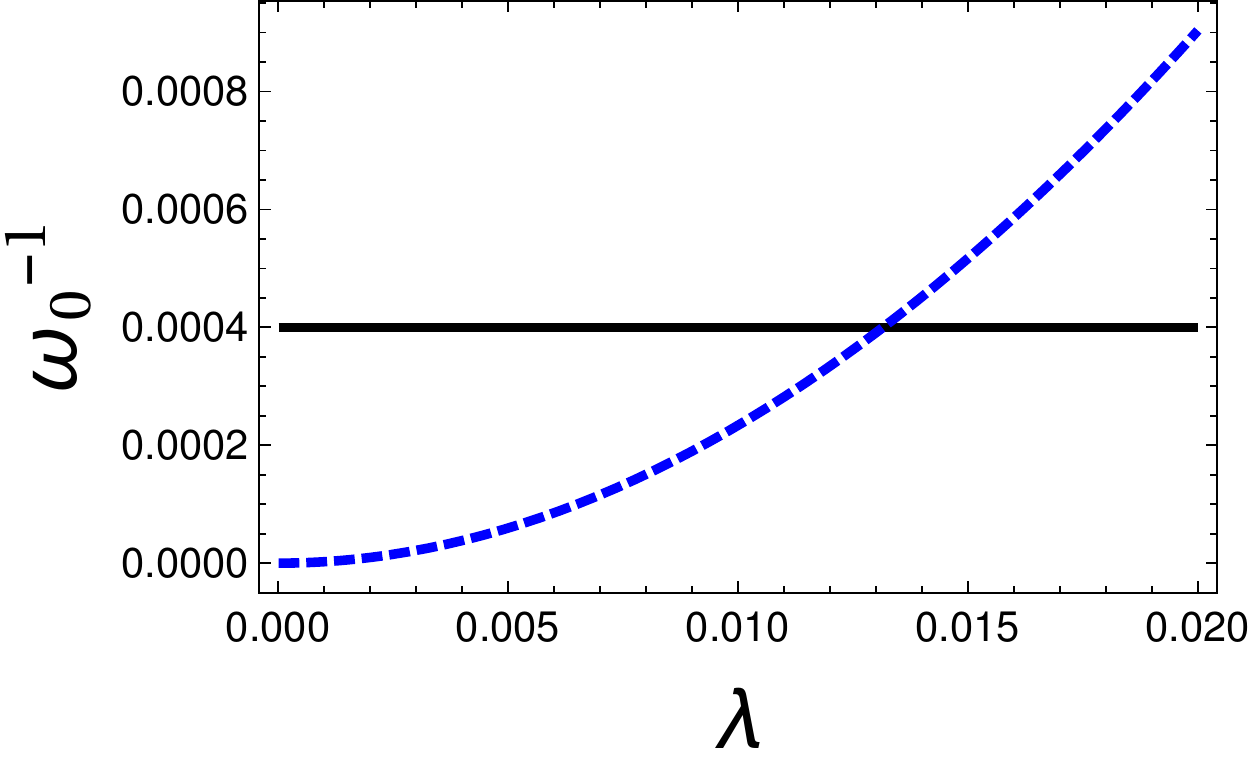}
\caption{\label{fig:5} 
Shown is the constraint from solar system tests for $p=1.25$ and $\Omega_{m,0}=0.27$.
}
\end{figure}

 As already mentioned, constraints from solar system tests require that \cite{leandros1}
\begin{equation}
\omega_0^{-1} = F_{,\phi}|_0^2 < 4 \times 10^{-4}
\end{equation}
Given that $F(\phi) = \lambda (\phi/\phi_0)^2$ and using eq. (15) we obtain the following expression for $\omega_0^{-1}$
\begin{equation}
\omega_0^{-1} = \frac{4 \lambda^2 (3p-2)^2}{8p \lambda-12 p^2 \Omega_{m,0}-4 \lambda (12p^2-11p+2)}
\end{equation}
where $\Omega_{m,0}$ is today's value of the normalized density of matter.
Fig. \ref{fig:5} shows $\omega_0^{-1}$ as a function of $\lambda$ for $\Omega_{m,0}=0.27$ and $p=1.25$. It is easy to check that the constraint from solar system tests requires that $\lambda < 0.013$. 


\subsection{Stability of cosmic structures}

\begin{figure}[ht!]
\centering
\includegraphics[width=\linewidth]{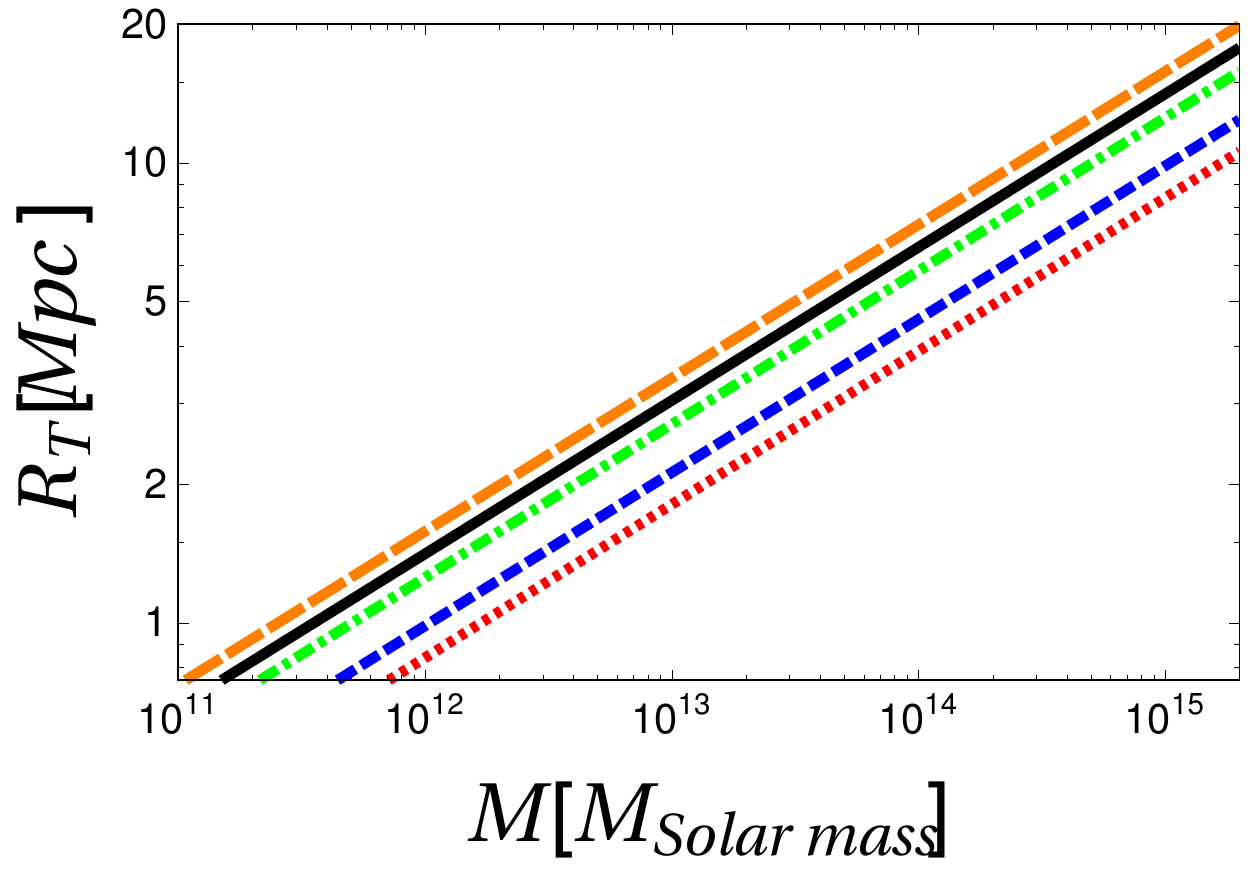}
\caption{\label{fig:6} 
Turnaround radius $R_T$ (in Mpc) versus mass $M$ (in solar masses)
for different values of the parameter $x=p/\lambda$ for $\Lambda$CDM (black), for dark energy with $w=-2.3$ (blue) and for $x=0.3,1,2$ from bottom to top.
}
\end{figure}

We recall that in 
\cite{tomaras1,tomaras2} it was shown that
in dark energy models with a constant equation-of-state parameter $w$ in GR, the stability of cosmic structures requires that $w > -2.3$.
In Fig. \ref{fig:6} we show the prediction for the maximum turnaround radius a) for the $\Lambda$CDM model (solid black line), b) for dark energy with $w=-2.3$ (short dashed blue line), c) for scalar-tensor cosmology for three different values of the parameter $x$, namely, $x=0.3$ (dotted red line), $x=1$ (dotted dashed green line) and $x=2$ (long dashed orange line). Therefore, our main result implies that the ratio $x=p/\lambda$ must satisfy the lower bound
\begin{equation}
x=\frac{p}{\lambda} > \frac{1}{2}
\end{equation}

\section{Conclusions}\label{Conclusions}

In this article we have analyzed a concrete model of Scalar-Tensor theory of gravity and we have obtained an exact power-law analytical solution (with matter included). Given that solution the expansion history as well as the effective Newton's constant can be computed explicitly, and the maximum turnaround radius can be computed in terms of the two free parameters of the model. Finally we have used several current observational data and constraints coming from supernovae, solar system tests and stability of cosmic structures to put bounds on the parameters of the model.

\section*{Acknowlegements}

We would like to thank T. N. Tomaras for spontaneous discussions that inspired our work, for reading the manuscript and for useful suggestions.
The author G. P. thanks the Funda\c c\~ao para a Ci\^encia e Tecnologia (FCT), Portugal, for the financial support to the Multidisciplinary 
Center for Astrophysics (CENTRA),  Instituto Superior T\'ecnico,  Universidade de Lisboa,  through the Grant No. UID/FIS/00099/2013. The author A.R. was supported by the CONICYT-PCHA/\- Doctorado Nacional/2015-21151658. 
We are grateful to L. Parivola\-ropoulos for emailing us the file with the Union 2 supernovae data.

\end{document}